\documentclass[preprintnumbers,superscriptaddress,amsmath,amssymb,prd,preprint,nofootinbib]{revtex4-2}

\usepackage{graphicx}
\usepackage{dcolumn}
\usepackage{bm}
\usepackage{color}
\usepackage{amssymb}
\usepackage{mathrsfs}
\usepackage{amsthm}
\usepackage{graphicx}
\usepackage{amsfonts}
\usepackage[colorlinks,linkcolor=red,anchorcolor=blue,citecolor=green]{hyperref}
\usepackage{subfigure}
\usepackage{float}
\usepackage{enumerate}

\begin{document}

%\preprint{APS/123-QED}

\title{Accretion of the relativistic Vlasov gas onto a Kerr black hole}

\author{Ping Li}
\email[]{Lip57120@huas.edu.cn}
\affiliation{College of Mathematics and Physics, Hunan University of Arts and Sciences, 3150 Dongting Dadao, Changde City, Hunan Province 415000, China}
\author{Yong-qiang Liu}
\email[]{1000511286@smail.shnu.edu.cn}
\affiliation{Division of Mathematica and Theoretical Physics, Shanghai Normal University, 100 Guilin Road, Shanghai 200234, China}
\author{Xiang-hua Zhai}
\email[]{zhaixh@shnu.edu.cn}
\affiliation{Division of Mathematica and Theoretical Physics, Shanghai Normal University, 100 Guilin Road, Shanghai 200234, China}

\date{\today}% It is always \today, today,

\begin{abstract}
We study the accretion of relativistic Vlasov gas onto a Kerr black hole, regarding the particles as distributed throughout all the space, other than just in the equatorial plane. We solve the relativistic Liouville equation in the full $3+1$ dimensional framework of Kerr geometry. For the flow that is stationary and axial symmetric, we prove that the distribution function is independent of the conjugate coordinates. For an explicit distribution that can approximate to Maxwell-J\"{u}ttner distribution, we further calculate the particle current density, the stress energy momentum tensor and the unit accretion rates of mass, energy and angular momentum.
	The analytic results at large distance are shown to be consistent with the limits of the numerical ones computed at finite distance. Especially, we show that the unit mass accretion rate agrees with the Schwarzschild result in the case of low temperature limit. Furthermore, we find from the numerical results that the
three unit accretion rates vary with  the angle in Kerr metric and the accretion of Vlasov gas would slow down the Kerr black hole. The closer to the equator, the faster it slows down the black hole.
\end{abstract}

\maketitle

%\tableofcontents

\section{Introduction}
The ensemble composed by a large number of massive particles, which interact with each other only through gravitation, is called Vlasov gas. The distribution function of such an ensemble satisfies collisionless Liouville equation, also called the Vlasov equation in mathematics. Vlasov models have a long history in astrophysics. The Newtonian systems are often used to model galaxies \cite{Fridman1984} and globular clusters \cite{Binney2008}. The relativistic Vlasov gas is often chosen to be the matter for investigating the open problems in general relativity, such as the cosmic censorship hypotheses\cite{cosmic}.
Especially, in the strong gravity areas, for example the galaxies' center, the interest of studying relativistic Vlasov gas is increasing \cite{Gunther2021-1,Gunther2021-2}.
For recent reviews, we refer to \cite{Rein2023}.

	It is claimed that the gas surrounding M87 or Sgr A*, both of which are the astronomical black holes observed by the Event Horizon Telescope Collaboration, is also nearly collisionless and magnetized\cite{Ryan2018,EHTL1,EHTL5,EHTL12,EHTL16}. In Refs. \cite{Rioseco2017-1,Rioseco2017-2}, Rioseco and Sarbach performed a detailed study of the accretion of Vlasov gas onto a Schwarzschild black hole. They solved the relativistic Liouville equation in Schwarzschild geometry and obtained the condition satisfied by the distribution function with an appropriate symmetry. In a particular model with a spherical steady-state flow, they derived the current density and the stress energy-momentum tensor. Such a relativistic Vlasov model is also used to provide a partial explanation for low accretion rate problem, also called the Bondi-Hoyle-Lyttleton (BHL) accretion problem\cite{Hoyle1939,Lyttleton1940,Bondi1944,Chaverra2015}.

	Since then, numerous works have been done, extending such a model to more general situations. Cie\'{s}lik and Mach generalized Rioseco and Sarbach's results to the Reissner-Nordstr\"{o}m black hole \cite{Cieslik2020}. In Refs. \cite{Mach2021-1,Mach2021-2,Mach2022}, authors calculated the accretion of relativistic Vlasov gas onto a moving Schwarzschild black hole and tried to resolve the BHL accretion problem. The accretion of Vlasov gas onto a Schwarzschild black hole at a sphere of finite radius is also studied by Gamboa \textit{et. al.} \cite{Gamboa2021}. Except for Schwarzschild black holes, Rioseco and Sarbach's results have also been extended to black holes in modified gravities. Liao and Liu investigated the accretion of the collisionless Vlasov gas onto a Bardeen regular black hole \cite{Liao2022}. Cai and Yang studied the accretion of Vlasov gas onto black holes in bumblebee gravity \cite{Cai2022}. There are also many researches studying the accretion onto Kerr black holes.
Ref. \cite{Rioseco2018} showed that a collisionless, relativistic kinetic gas configuration propagating in the equatorial plane of a Kerr black hole would eventually settle down to a stationary, axisymmetric configuration surrounding the black hole. In the slowly rotating case, Ref. \cite{Andersson2018} proved the decay of a bounded energy and integrated energy for massless Vlasov gas in Kerr spacetime. More recently, Cie\'{s}lik \textit{et. al.} performed a detailed study of the accretion of Vlasov gas onto a Kerr black hole, occurring in the equatorial plane \cite{Cieslik2022}. In Ref. \cite{Mach2023}, Mach et.al. studied the equatorial accretion on a moving Kerr black hole.

	However, the studies about Vlasov gas accreting onto Kerr black holes so far have utilized the result of the distribution function of Vlasov gas obtained in Ref. \cite{Rioseco2017-1}, which is one of the solutions of the collisionless Liouville equation in Schwarzschild geometry. When considering the accretion onto a Kerr black hole, it is reasonable that the distribution function is the solution of the Liouville equation in Kerr geometry. In this paper, we solve the relativistic Liouville equation in the full 3+1 dimensional frame of Kerr geometry and show that the general solution of Liouville equation in Kerr geometry is different from the axisymmetric case in Schwarzschild geometry.
On the other hand, since the total angular momentum $L^2$ is conserved in Schwarzschild geometry, it is reasonable to consider the time-like geodesic bounded to a single plane. However, things become different when working in Kerr geometry where the conserved quantity turns into the Carter constant $D$ but no longer the total angular momentum $L^2$. It is not enough to work in 2+1 framework when talking about the accretion of Vlasov gas in Kerr geometry. In this paper, we discuss this issue in a full 3+1 framework. Then the particle current density $J_{\mu}$ and the stress energy-momentum tensor $T_{\mu\nu}$ depend on both radius $r$ and angular $\theta$. Therefore, when talking about the accretion of Vlasov gas in Kerr geometry, it is more complete to work in a full 3+1 framework than to consider only the equatorial plane.

	This paper is organized as follows. In Sec.II, we review the Hamiltonian description of time-like geodesics in Kerr geometry. In Sec.III, we solve the relativistic Liouville equation in a Kerr background and prove that the distribution function only depends on canonical momentum $P_{\mu}$. Furthermore, we simplify the volume element in both non-equatorial plane and equatorial plane. In the special case that $2$-surface is a spherical surface, we obtain the expression of three accretion rates per unit surface. In Sec.IV, we consider an explicit collisionless flow in stationary state and with axial symmetry  that can approximate to Maxwell-J\"{u}ttner distribution at infinity. The expression of the particle current density $J_{\mu}$ and the stress energy-momentum tensor $T_{\mu\nu}$ are computed. The limits of integration is discussed. In Sec.V, we analytically examine the asymptotic behavior in large distances where the corresponding quantities are all independent of angular $\theta$. Using Taylor expansion of $\frac{1}{r}$, the leading term and second term of unit accretion rates are computed. In Sec.VI, we numerically compute the corresponding quantities in finite ranges. The last section is a detailed conclusion. In this paper, we use the units $G=c=1$.

\section{Review on geodesic motion in Kerr spacetime}
\subsection{Hamilton formulas of geodesic motion in Kerr spacetime}

	The geodesic motion in Kerr spacetime has been widely discussed in many papers and books. In this section, we give a brief review. This is done in Boyer-Lindquist coordinates $(t,r,\theta,\varphi)$, where the Kerr metric can be written as
\begin{align}\label{Kerr}
ds^2=\left(1-\frac{2Mr}{\rho^2}\right)dt^2+\frac{4Mar \sin^2\theta}{\rho^2}dtd\varphi
  -\frac{\rho^2}{\Delta}dr^2 -\rho^2d\theta^2-\frac{\sin^2\theta}{\rho^2}\Sigma^2d\varphi^2,
\end{align}
where
\begin{align}
  \rho^2 &=r^2+a^2\cos^2\theta,  \\
  \Delta &=r^2-2Mr+a^2,\\
  \Sigma^2&=(r^2+a^2)^2-a^2\Delta\sin^2\theta.
\end{align}
The inverse Kerr metric can be expressed as
\begin{equation}
  [g^{\mu\nu}]=\left(
                 \begin{array}{cccc}
                   \frac{\Sigma^2}{\Delta \rho^2} & 0 & 0 & \frac{2Mar}{\Delta \rho^2} \\
                   0 & -\frac{\Delta}{\rho^2} & 0 & 0 \\
                   0 & 0 &  -\frac{1}{\rho^2} & 0 \\
                   \frac{2Mar}{\Delta \rho^2} & 0 & 0 & -\frac{\Delta-a^2\sin^2\theta}{\rho^2\Delta\sin^2\theta} \\
                 \end{array}
               \right),
\end{equation}
which satisfies $g^{\mu\alpha}g_{\alpha\nu}=\delta^{\mu}_{\nu}$. This metric (\ref{Kerr}) is characterized by the black hole's mass $M$ and  angular momentum	 $J=Ma$. In this paper, we only concern the range $r>r_{+}$, where $r_{+}=M+\sqrt{M^2-a^2}$ is the outer horizon.

	The geodesic motion in Kerr spacetime can be written as decoupled, first-order equations
\begin{align}
  \rho^2\dot{t} & =\frac{1}{\Delta}\left(\Sigma^2 E-2aMrL_z \right), \\
  \rho^2\dot{\varphi} &=\frac{1}{\Delta}\left(2aMrE+(\rho^2-2Mr)L_z\csc^2\theta\right),\\
   \rho^4\dot{r}^2&\equiv  R =E^2r^4+(a^2E^2-L_z^2-D)r^2-a^2D+2Mr(D+(L_z-aE)^2)-m^2r\Delta,\label{R}\\
   \rho^4\dot{\theta}^2&\equiv \Theta= D+(a^2E^2-L_z^2\csc^2\theta)\cos^2\theta -m^2a^2\cos^2\theta,\label{sita}
\end{align}
where the dot stands for the derivative with respect to the proper time $s$. These equations involve four constants of motion: the rest mass $m$, the energy $E$, the angular momentum $L_z$ in $z$ direction and the Carter constant $D$. The rest mass $m$ is defined by
\begin{equation}
  m^2=g_{\mu\nu}\dot{x}^{\mu}\dot{x}^{\nu}.
\end{equation}
Equations with $m=0$ stand for null geodesic, and with $m\neq0$ stand for time-like geodesic. In this paper, we only concern the case of $m\neq0$. The energy $E$ and the $z$-axis angular momentum $L_z$ are defined by
\begin{align}
 E & = \left(1-\frac{2Mr}{\rho^2}\right)\dot{t}+\frac{2Mar \sin^2\theta}{\rho^2}\dot{\varphi},\\
  L_z & =-\frac{2Mar \sin^2\theta}{\rho^2}\dot{t}+ \frac{\sin^2\theta}{\rho^2}\Sigma^2\dot{\varphi}.
\end{align}
The Carter constant D is defined by
\begin{equation}
  D=\frac{1}{\Delta}(\Delta \dot{t}-a\Delta\sin^2\theta\dot{\varphi})^2-\frac{\rho^4}{\Delta}\dot{r}^2-m^2r^2-(L_z-aE)^2.
\end{equation}
Because the angular momentum $L^2$ is no longer a conserved quantity, the geodesic cannot travel in the same plane. This is the main difference of geodesic motion between in Kerr background and in Schwarzschild background.

	The Hamilton formulas can be used to describe the particle motion in Kerr spacetime. The Hamiltonian $H=\frac{1}{2}g_{\mu \nu}p^{\mu} p^{\nu}=\frac{m^{2}}{2}$ of geodesic motion is expressed as
\begin{widetext}
\begin{equation}
  H=\frac{1}{2}\left(\left(1-\frac{2Mr}{\rho^2}\right)\dot{t}^2+\frac{4Mar \sin^2\theta}{\rho^2}\dot{t}\dot{\varphi}
  -\frac{\rho^2}{\Delta}\dot{r}^2-\rho^2\dot{\theta}^2-\frac{\sin^2\theta}{\rho^2}\Sigma^2\dot{\varphi}^2 \right).
\end{equation}
\end{widetext}
The corresponding canonical momentums $(p_t,p_r,p_{\theta},p_{\varphi})$ can be calculated by
\begin{align}
  p_t&=\frac{\partial H}{\partial \dot{t}}=E, \\
  p_{\varphi}&=-\frac{\partial H}{\partial \dot{\varphi}}=-L_z,\\
  p_r&=-\frac{\partial H}{\partial \dot{r}}=\frac{\rho^2}{\Delta}\dot{r}=\pm\frac{\sqrt{R}}{\Delta},\\
  p_{\theta}&=-\frac{\partial H}{\partial \dot{\theta}}=\rho^2\dot{\theta}=\pm\sqrt{\Theta}.
\end{align}
Since the Hamiltonian does not contain $t$ and $\varphi$, the Hamilton formulas show that $E$ and $L_z$ are constants, which are
\begin{equation}
  \dot{p}_t=\frac{\partial H}{\partial t}=0,~~~~~~~\dot{p}_{\varphi}=\frac{\partial H}{\partial \varphi}=0.\nonumber
\end{equation}
The symplectic variables $(x^{\mu},p_{\mu})$ are understood as phase space coordinates.
\subsection{The $r$-motion}
As $\rho^4 \dot{r}^2=R $, function $\frac{R}{\rho^4}$ can be understood as the equivalent kinetic energy in radial direction, whose properties are mainly determined by function $R$. Simplifying equation (\ref{R}), we obtain
\begin{align}
  R=& (E^2-m^2)r^4+2m^2Mr^3+(a^2(E^2-m^2)-L_z^2-D)r^2+2M(D+(L_z-aE)^2)r-a^2D.\label{Rr}
\end{align}
Particles in Kerr background appear with non-negative kinetic energy, implying $R\geq 0$. Notice that  $R$ only depends on $r$ but not on $\theta$, the properties of $r$-motion are irrelevant to $\theta$.

	We consider the particles incident from infinity towards the black hole. The requirement $R\geq 0 (r\rightarrow\infty)$ indicates that the energy satisfies $E\geq m$. Generally speaking, function $R(r)$ has four zero points including both real and imaginary roots. For physical reasons, we only discuss positive real points. Let's define $r_0$ as the biggest real root of the equation $R=0$. A test particle coming from infinity can travel to $r=r_0$. Once arriving at $r=r_0$, its radial kinetic energy becomes zero. There exist three cases to discuss.
\begin{enumerate}
  \item $r_0<r_+$: The test particle falls into the black hole.
  \item $r_0>r_+$, and $R<0$ for $r<r_0$: The test particle cannot enter the range $r<r_0$. It will be scattered by the black hole and travel to infinity. In this case, the position $r=r_0$ is the closest location, also known as the perihelion $r_p=r_0$, when particle traveling through the black hole.
  \item $r_0>r_+$, and $R>0$ for $r<r_0$: The test particle cannot travel to $r<r_0$ but keep getting close to $r=r_0$ and never reach there. The position $r=r_0$ stands for an unstable orbit $r_c=r_0$.
\end{enumerate}
Case 3 is a critical situation lying between case 1 and case 2. Since $R(r)>0$ is true for both $r>r_0$ and $r<r_0$, $R(r_0)=0$ is the minimum of function $R(r)$. Then, case 3 satisfies
\begin{align}
  R(r_c,D_c) &=0,\label{eq1}  \\
 \partial_{r}R(r_c,D_c) & =0.\label{eq2}
\end{align}

\begin{figure*}
\centering
\includegraphics[width=.45\textwidth]{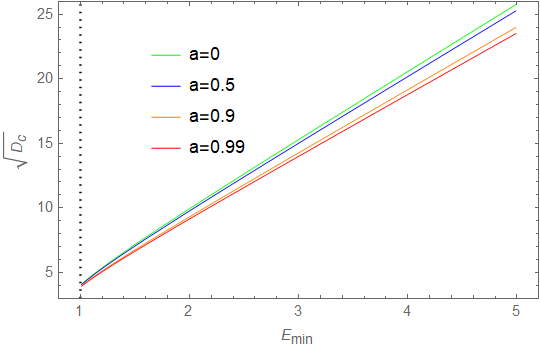}
\includegraphics[width=.45\textwidth]{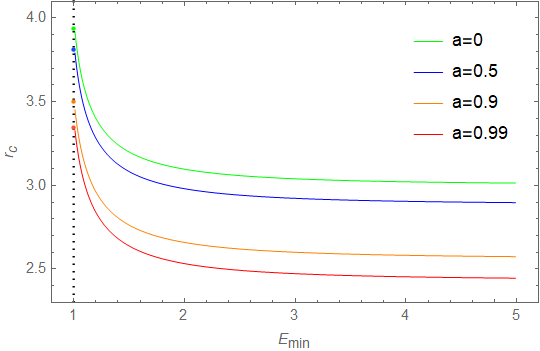}
\caption{The critical parameter $\sqrt{D_c}$ (left) and the unstable orbit $r_c$ (right) as functions of the energy $E_{min}$. Other parameters are chosen to be $M=1,m=0.01,L_z=0$.}
\label{Fig:Dcrc}
\end{figure*}

	The critical Carter constant parameter $D_c=D_c(E_{min},L_z,m)$ determines whether there is a perihelion along the geodesic motion, where $E_{min}$ is the minimum energy possessed by a particle that can escape from the black hole. For geodesic motion with the parameter $D>D_c$, the perihelion exists and the particle should be scattered to infinity by the black hole; Otherwise, the perihelion does not exist and the particle should be captured by the black hole.
%Notice that $D_c(E_{min},L_z,m)$ is actually determined by the energy $E_{min}$, the angular momentum $L_z$ and the mass $m$.
	%Since $R(r)$ is a biquadratic function, the analytic process of obtaining zero points is difficult. In this paper, we use numerical method to do the calculation.
	The numerical results are shown in Fig.\ref{Fig:Dcrc}. The perihelion $r_p$ only appears in the geodesic motion with parameter $D>D_c$. Besides, the perihelion $r_p$ lies in the range $r_p>r_c$.
	
%{\begin{figure}
%\centering
%\includegraphics[width=.45\textwidth]{E-Dc.png}
%\caption{The critical parameter $\sqrt{D_c}$ as functions of the energy $E_{min}$. Other parameters are chosen to be $M=1,m=0.01,L_z=0$.}
%\label{Fig:Dcrc1}
%\end{figure}

%\begin{figure}
%\centering
%\includegraphics[width=.45\textwidth]{E-rc.png}
%\caption{The unstable orbit $r_c$ as functions of the energy $E_{min}$. Other parameters are chosen to be $M=1,m=0.01,L_z=0$.}
%\label{Fig:Dcrc}
%\end{figure}

\subsection{The $\theta$-motion}

	The behavior of $\theta$-motion is similar to the $r$-motion. As $\rho^4 \dot{\theta}^2=\Theta$, we take $\frac{\Theta}{\rho^4}$ as the equivalent kinetic energy in $\theta$ direction, which is mainly influenced by the function $\Theta(\theta)$. Defining $y=\cos^2\theta\in[0,1]$, one simplifies Eq. (\ref{sita}) to
\begin{equation}
  \Theta=D+\left(a^2(E^2-m^2)-\frac{L_z^2}{1-y} \right)y.
\end{equation}
The root $y_0$ of $\Theta|_{y=y_0}=0$ determines the range of $\theta$ in geodesic motion. We only concern the physical situation that the zero point $y_0$ lies in $y_0\in[0,1]$. It is further divided into two cases.
\begin{enumerate}
  \item $L_z=0$: One obtains
  \begin{equation}
    \Theta=D+a^2(E^2-m^2) y,
  \end{equation}
which is monotonically increasing if $E>m$. For the parameter $D\geq0$, the result $\Theta|_{y=0} \cdot \Theta|_{y=1}=D(D+a^2(E^2-m^2))>0$ indicates no zero points exist in the range $[0,1]$. That is to say, the kinetic energy in $\theta$ direction is always positive and the test particle can travel throughout the range of $\theta\in[0,\pi]$.
  \item $L_z\neq 0$: It can be proven that there is only one zero point in the interval $[0,1]$. Denoting $y=y_0$ is the root, then the particle can move in the range $\theta\in[\theta_0,\pi-\theta_0]$, where $\theta_0=\arccos\sqrt{y_0}\in[0,\frac{\pi}{2}]$.
\end{enumerate}

\section{Vlasov gas in Kerr geometry}
\subsection{The distribution function of Vlasov gas}
	
	The Vlasov gas is considered as a collection of relativistic, free and collisionless kinetic particles moving along the geodesic lines. Unlike the accretion onto a Schwarzschild black hole, the accretion onto a Kerr black hole is no longer spherically symmetric, and considering only in one plane is not enough. The distribution function $f(x^{\mu},p_{\mu})$ satisfies the collisionless Liouville equation
\begin{equation}\label{Boltzmann}
  \dot{f}\equiv \{f,H\}=0,
\end{equation}
where the bracket $\{.,.\}$ stands for the Poisson bracket. The spherically symmetric solution of Eq. (\ref{Boltzmann}) has been discussed detailedly in Ref. \cite{Rioseco2017-1}. Similar to this, we seek the axisymmetric solution based on the transformation of symplectic variables $(x^{\mu},p_{\mu})\rightarrow (Q^{\mu},P_{\mu})$.

	Following Carter's study \cite{Carter}, the abbreviated action $S$ in Kerr geometry can be written as
\begin{equation}
  S=\frac{1}{2}m^2\tau-Et+L_z\varphi +\int_{r}\frac{\sqrt{R}}{\Delta}dr+\int_{\theta}\sqrt{\Theta}d\theta,
\end{equation}
where the integral bounds are the intervals of the geodesic motion. The new momenta $P_{\mu}$ are defined by the conserved quantities
\begin{align}\label{NewM}
  P_0 & =\sqrt{2H}= m, \\
 P_1 & =E,\\
 P_2&=-L_z,\\
 P_3&=\sqrt{D}.
\end{align}
The corresponding conjugate coordinates $Q^{\mu}$ are obtained by the following calculation
\begin{align}
  Q^0 &=\frac{\partial S}{\partial m}=m\left(\tau-\int_r\frac{r^2}{\sqrt{R}}dr-\int_{\theta}\frac{a^2\cos^2\theta}{\sqrt{\Theta}}d\theta\right),  \\
  Q^1 &= \frac{\partial S}{\partial E}=-t+\int_r\frac{Er^4+a^2Er^2-2Mar(L_z-aE)}{\Delta\sqrt{R}}dr+\int_{\theta}\frac{a^2E\cos^2\theta}{\sqrt{\Theta}}d\theta,\\
  Q^2 &= \frac{\partial S}{\partial (-L_z)}=-\varphi+\int_r\frac{L_z r^2-2(L_z-aE)Mr}{\Delta\sqrt{R}}dr +\int_{\theta}\frac{L_z\cot^2\theta}{\sqrt{\Theta}}d\theta,\\
  Q^3&=\frac{\partial S}{\partial \sqrt{D}}=\sqrt{D}\left(-\int_r\frac{1}{\sqrt{R}}dr+
  \int_{\theta}\frac{1}{\sqrt{\Theta}}d\theta\right).
\end{align}

	The canonical transformations $(x^{\mu},p_{\mu})\rightarrow (Q^{\mu},P_{\mu})$ keep the Poisson bracket (\ref{Boltzmann}) covariant
\begin{equation}
  \frac{\partial H}{\partial p_{\mu}}\frac{\partial}{\partial x^{\mu}}-\frac{\partial H}{\partial x^{\mu}}\frac{\partial}{\partial p_{\mu}}
  =  \frac{\partial H}{\partial P_{\mu}}\frac{\partial}{\partial Q^{\mu}}-\frac{\partial H}{\partial Q^{\mu}}\frac{\partial}{\partial P_{\mu}}.
\end{equation}
In the new canonical variables $(Q^{\mu},P_{\mu})$, the Hamiltonian becomes $H=\frac{P_0^2}{2}$. Therefore, the collisionless Liouville equation (\ref{Boltzmann}) is simplified to
\begin{equation}
  \frac{\partial}{\partial Q^0}f=0.
\end{equation}
The general solution is given by
\begin{equation}
 f= f(Q^1,Q^2,Q^3,P_0,P_1,P_2,P_3).
\end{equation}
Besides, there exist more symmetry conditions to restrict the distribution function (\ref{Boltzmann}) from Kerr geometry. The Kerr metric is independent of $t$ and $\varphi$ which appear in $Q^1$ and $Q^2$. Then, the distribution function $f$ can be simplified to a more simple form $f= f(Q^3,P_0,P_1,P_2,P_3)$. Moreover, as $D$ is a constant, one has $\frac{\partial S}{\partial D}=0$ \cite{Chandrasekhar1983} that further leads to $Q^3=0$. At last, it is indicated that the distribution function $f$ in Kerr geometry depends only on the new momenta $P_{\mu}$
\begin{equation}
   f= f(P_0,P_1,P_2,P_3).\label{condition}
\end{equation}
\subsection{Observable Quantities}
The observable quantities we concern are the particle current density $J_{\mu}$ and the stress energy-momentum tensor $T_{\mu\nu}$, which are defined at spacetime point $x$ of Kerr spacetime $\mathscr{M}$ by
\begin{align}
  J_{\mu}|_{x} &= \int_{\pi} p_{\mu}f(x^{\gamma},p_{\gamma})\sqrt{-\det (g^{\alpha\beta})}d^4p,\label{J} \\
  T_{\mu\nu}|_{x} & =\int_{\pi} p_{\mu}p_{\nu}f(x^{\gamma},p_{\gamma})\sqrt{-\det (g^{\alpha\beta})}d^4p,\label{T}
\end{align}
where the integral range $\pi$ is the subset of the cotangent space $T^{*}_{x}\mathscr{M}$, which will be determined  according to the properties of geodesic. Using Eq. (\ref{Boltzmann}), one can show that the particle current density $J_{\mu}$ and the energy-momentum tensor $T_{\mu\nu}$ satisfy the conservation laws
\begin{align}
  \nabla_{\mu}J^{\mu} &=0,  \\
   \nabla_{\mu}T^{\mu\nu}& =0.
\end{align}
The particle number density $n_s$ is defined by $J_{\mu}$ through the relationship
\begin{equation}\label{number}
  n_s=\sqrt{g^{\mu\nu}J_{\mu}J_{\nu}}.
\end{equation}

	In the actual calculation, the momentum variables $(p_t,p_r,p_{\theta},p_{\varphi})$ transform to variables $(m^2,E,L_z,D)$, which can be re-expressed as follows
\begin{align}
  m^2 &=g^{00}p_t^2+2g^{03}p_tp_{\varphi}+g^{33}p_{\varphi}^2+g^{11}p_r^2+g^{22}p_{\theta}^2,  \\
   E& =p_t,\\
   K&=\frac{(r^2+a^2)^2}{\Delta}p_t^2+\frac{2a(a^2+r^2)}{\Delta}p_tp_\varphi+
   \frac{a^2}{\Delta}p_{\varphi}^2-\Delta p_r^2-m^2r^2,\\
   L_z&=-p_{\varphi}.
\end{align}
Using above equations, we obtain the Jacobian determinant
\begin{equation}
  J=\frac{\partial(m^2,E,L_z,D)}{\partial(p_t,p_r,p_{\theta},p_{\varphi})}=\frac{4\sqrt{R\Theta}}{\rho^2},
\end{equation}
and simplify the volume element as
\begin{equation}
  \sqrt{-\det(g^{\mu\nu})}dp_tdp_rdp_{\theta}dp_{\varphi}=\frac{dm^2dEdL_zdD}{4\sqrt{R\Theta}\sin\theta}.
\end{equation}
Next, we will further simplify the volume element in two different cases.

~\

{\noindent} 1 Non-equatorial plane $\theta\neq\frac{\pi}{2}$

~\

	In non-equatorial plane $\theta\neq\frac{\pi}{2}$, there exist constraints on $L_z$ to ensure $\Theta>0$. Let's define
\begin{align}
 X^2&=\tau\sin^2\theta+(E^2-m^2)a^2\sin^2\theta, \label{X}\\
\tau&=\frac{D}{\cos^2\theta}.
\end{align}
Inserting these expressions into Eq. (\ref{sita}), one obtains
\begin{equation}
  \Theta \sin^2\theta=\cos^2\theta(X^2-L_z^2).
\end{equation}
Thus, the integral range of $L_z$ is $L_z\in[-X,X]$. Furthermore, we make a substitution
\begin{equation}
  L_z=X\sin\sigma,
\end{equation}
and obtain
\begin{equation}
  \frac{dL_z}{\sqrt{\Theta \sin^2\theta}}=\frac{d\sigma}{|\cos\theta|}.
\end{equation}
The integral range of $\sigma$ is $\sigma\in[-\frac{\pi}{2},\frac{\pi}{2}]$. At last, the volume element is re-expressed as
\begin{equation}
  \sqrt{-\det(g^{\mu\nu})}dp_tdp_rdp_{\theta}dp_{\varphi}=\frac{|\cos\theta| }{4\sqrt{R}}dm^2dE d\tau d\sigma.
\end{equation}

Notice that the definition (50) becomes divergent when $\theta=\pi/2$, thus the above expression for the volume element is not applicable to the case of equatorial plane. Therefore, for the case of $\theta=\pi/2$ we have to do the calculations separately.

~\

{\noindent}2 Equatorial plane $\theta=\frac{\pi}{2}$

~\

	In the equatorial plane $\theta=\frac{\pi}{2}$, $\Theta$ degenerates to $D$. There is no constraint on $L_z$. The integral range of $L_z$ is simply $L_z\in(-\infty,\infty)$ and the volume element is expressed as
\begin{equation}
  \sqrt{-\det(g^{\mu\nu})}dp_tdp_rdp_{\theta}dp_{\varphi}=\frac{dm^2dEdL_zdD}{4\sqrt{D}\sqrt{R}}.
\end{equation}

		In section II, the geodesic motion coming from infinity, also named unbounded orbit and satisfying $E>m$, was divided to three classifications. The key of the classification is based on the values of parameters $(m^2,E,L_z,D)$ or equivalently parameters $(m^2,E,\sigma,\tau)$. For particles absorbed by the black hole, there is no perihelion along their trajectories, which indicates that the parameter $\tau$ lies in the range $\tau<\tau_c$. For particles scattered by the black hole, the perihelion $r_p$ exits and the parameter $\tau$ is in the range $\tau>\tau_c$. In the critical case between them, particles are neither absorbed nor scattered by the black hole, but get infinitely close to the unstable orbit $r_c$. In this case, the value of parameter $\tau$ is $\tau=\tau_c$. According to the classification of orbit, the observable quantity $J_{\mu}$ or $T_{\mu\nu}$ consists of three parts: the absorbed part, the scattered part and the critical part. That is,
\begin{align}
  J_{\mu} &=J^{abs}_{\mu}+J^{scat}_{\mu}+J^{cri}_{\mu},  \\
  T_{\mu\nu} &=T^{abs}_{\mu\nu}+T^{scat}_{\mu\nu}+T^{cri}_{\mu\nu},
\end{align}
where the critical part does not have substantial contribution and attract no attention in most cases. Notice that the observable quantities (\ref{J}) and (\ref{T}) depend on both $r$ and $\theta$. If the distribution function $f(x^{\mu},p_{\mu})$ is given, $J_{\mu}$ and $T_{\mu\nu}$ can be calculated.

\subsection{The accretion rates}

	Because the geodesic motion is not in a single plane, the collisionless particles are no longer in a single plane, too. We suppose that they are distributed throughout all the space outside the black hole. Since the flow is stationary and axisymmetric, the equation $\nabla_{\mu} J^{\mu}=0$ becomes
\begin{equation}
  (\rho^2\sin\theta J^{r})_{,r}+(\rho^2\sin\theta J^{\theta})_{,\theta}=0.
\end{equation}
Let $V$ be a region in Kerr spacetime and $S$ be a 2-surface boundary of $V$. The above equation shows that the integral of left side in $V$ is a constant, which is called the mass accretion rate. Consider vector $\hat{n}=(n_r,n_{\theta},0)$ is the unit normal field of $S$ that directs outside. Using the Stokes theorem, the mass accretion rate is expressed as
\begin{align}
\frac{d M}{dt}&=-\int_V\left((\rho^2\sin\theta J^{r})_{,r}+(\rho^2\sin\theta J^{\theta})_{,\theta}\right) dV,\\
&=-\int_S( J^{r}\hat{n}_r+J^{\theta} \hat{n}_{\theta})\rho^2 d\Omega,
\end{align}
where $d\Omega=\sin\theta d\theta d\varphi$ and the current flux $J^{\mu}$ traverse through $S$. On the other hand, $T_{\mu\nu}$ are also divergence-free, so there exist two more accretion rates besides mass accretion rate in Kerr geometry. They are energy and angular momentum accretion rates as follows
\begin{align}
\frac{d \mathcal{E}}{dt} &= -\int_S(T^r{}_{t}\hat{n}_r+T^{\theta}{}_{t}\hat{n}_{\theta})\rho^2 d\Omega,\\
  \frac{d\mathcal{L}}{dt}&=-\int_S(T^r{}_{\varphi}\hat{n}_r+T^{\theta}{}_{\varphi}\hat{n}_{\theta})\rho^2 d\Omega.
\end{align}

	As considered by other authors, we only concern that 2-surface $S$ is a spherical surface with the surface element $dS=r^2d\Omega$, where the unit external normal vector is chosen by $\hat{n}=(1,0,0)$. The corresponding accretion rates through a unit surface are expressed as
\begin{align}
 \frac{d^2 M}{dSdt}&=J_r\frac{\Delta}{r^2} ,\label{a1}\\
  \frac{d^2 \mathcal{E}}{dSdt} &= T_r{}_{t}\frac{\Delta}{r^2} ,\label{a2}\\
  \frac{d^2\mathcal{L}}{dSdt}&=T_r{}_{\varphi}\frac{\Delta}{r^2}.\label{a3}
\end{align}

	\section{explicit example and limits of integration}
We consider a stationary, axially symmetric, collisionless gas distributed throughout all of the Kerr spacetime. Notice that it is no longer a disk confined to a constant $\theta_0$ plane. At infinity, Kerr metric is asymptotically flat and we suppose the gas satisfies the same asymptotic condition as in spherically symmetric spacetime. As an explicit example, we choose the distribution function
\begin{align}
  f&=\delta(P_0-m_0)f_{\infty}(P_1)|_{P_1=E} \nonumber\\
  &=A\delta(P_0-m_0)e^{-\beta P_1},\label{distribution}
\end{align}
where $m_0$ is the mass of particle and $A$ is the normalization factor. Variable $\beta=\frac{1}{k_bT}$ where $k_b$ is the Boltzmann constant and $T$ is the asymptotic temperature. The distribution function (\ref{distribution}) fulfills the conclusion (\ref{condition}) of Vlasov gas in Kerr geometry.
\subsection{Non-equatorial plane $\theta\neq\frac{\pi}{2}$}
Using dimensionless variables, we re-define
\begin{align}
  E&=m_0\varepsilon,\\
  \tau & =m_0^2\bar{\tau},\\
  X^2&=m_0^2 \bar{X}^2,\\
  R &=m_0^2\bar{R},
\end{align}
where $X$ is expressed as Eq. (\ref{X}) and $R$ is expressed as Eq. (\ref{Rr}). Therefore, we have
\begin{align}
  \bar{X}^2&= \left(\bar{\tau}+(\varepsilon^2-1) a^2 \right)\sin^2\theta,\label{X1}\\
  \Theta&=m_0^2\bar{X}^2\cos^2\sigma\frac{\cos^2\theta}{\sin^2\theta},\label{theta1}\\
 \bar{R}&=(\varepsilon^2-1)r^4+2M r^3+ (a^2(\varepsilon^2-1)-\bar{X}^2\sin^2\sigma-\bar{\tau}\cos^2\theta)r^2\notag\\
 & \quad+2M(\bar{\tau}\cos^2\theta+(\bar{X}\sin\sigma-a\varepsilon)^2)r-a^2\bar{\tau}\cos^2\theta.\label{R1}
\end{align}
Next, we will discuss the upper and lower limits of the integral element corresponding to the absorption part and the scattering part.

We solve the simultaneous equations
\begin{align}
   \partial_{r} \bar{R}&=0, \label{74} \\
   \bar{R}&=0, \label{75}
\end{align}
and obtain the solutions
\begin{align}
  r & =r_c(\varepsilon,\sigma;\theta),\\
  \bar{\tau} &=\bar{\tau}_c(\varepsilon,\sigma;\theta).
\end{align}
For the absorption part, as we have stated, there is no perihelion on these geodesic orbits. Thus, the interval of element $\bar{\tau}$ is in the range $\bar{\tau}\in[0,\bar{\tau}_c]$. From Eq. (\ref{theta1}), $\Theta$ is always positive, and no additional conditions are provided for constraining $\sigma$ from $\Theta$. So the range of $\sigma$ is $\sigma\in[-\frac{\pi}{2},\frac{\pi}{2}]$. The limits of element $\varepsilon$ are related to the energy condition of Vlasov gas. In this paper, we only concern that the black hole accretes particles travelling from infinity and the parameter $\varepsilon$ satisfies $\varepsilon\in[1,\infty)$. Plugging the distribution function (\ref{distribution}) into the observable quantities (\ref{J}) and (\ref{T}), we obtain the absorption part as follows
\begin{align}
  J^{abs}_t (r,\theta)&=Am_0^4\int_{-\frac{\pi}{2}}^{\frac{\pi}{2}}d\sigma\int_1^{\infty}d\varepsilon\int_{0}^{\bar{\tau}_c}d\bar{\tau}\frac{\varepsilon e^{-\bar{\beta}\varepsilon}|\cos\theta|}{2\sqrt{\bar{R}}},\label{Jt1}  \\
  J^{abs}_r(r,\theta) &=Am_0^4\int_{-\frac{\pi}{2}}^{\frac{\pi}{2}}d\sigma\int_1^{\infty}d\varepsilon\int_{0}^{\bar{\tau}_c}d\bar{\tau}\frac{ e^{-\bar{\beta}\varepsilon}|\cos\theta|}{2\Delta},\label{Jr1}\\
  J^{abs}_{\theta}(r,\theta)&=Am_0^4\int_{-\frac{\pi}{2}}^{\frac{\pi}{2}}d\sigma\int_1^{\infty}d\varepsilon\int_{0}^{\bar{\tau}_c}d\bar{\tau}\frac{\bar{X} e^{-\bar{\beta}\varepsilon}\cos\sigma\cos^2\theta}{2\sqrt{\bar{R}}\sin\theta},\label{Jtheta1}\\
  J^{abs}_{\varphi}(r,\theta)&=-Am_0^4\int_{-\frac{\pi}{2}}^{\frac{\pi}{2}}d\sigma\int_1^{\infty}d\varepsilon\int_{0}^{\bar{\tau}_c}d\bar{\tau}\frac{\bar{X} e^{-\bar{\beta}\varepsilon}\sin\sigma|\cos\theta|}{2\sqrt{\bar{R}}},\label{Jphi1}
\end{align}
where $\bar{\beta}=m_0\beta$. Notice that $T_{\mu\nu}$ have 10 components and we list them in appendix A.

By evaluating the integrals through new coordinates $x=\sqrt{\varepsilon^2-1}\sin\sigma, y=\sqrt{\varepsilon^2-1}\cos\sigma$, and $z=\frac{\bar{\tau}}{\varepsilon^2-1}$, we can prove that $J_{\theta}=0, T_{r\theta}=0, T_{\theta\phi}=0, T_{t\theta}=0$ and $J_{\phi} \neq 0, T_{t\phi}\neq 0$. Since we consider the fluid passing through a 2-sphere, and as is shown in Eq. (58), the accretion rates are defined by two Killing vectors $\partial_t$ and $\partial_{\phi}$. Thus, $J^{\phi}$ and $T^t{}_{\phi}$ would not appear in the accretion rates. On the other hand, to compare with the accretion of an isotropic perfect fluid is also an interesting topic [13] but it is not within the scope of this paper. We will do a more detailed study in the future to compare our model with the accretion of a perfect fluid.

	For the scattering part, the range of $\sigma$ is also simply in $\sigma\in[-\frac{\pi}{2},\frac{\pi}{2}]$.  Particles with energy $\varepsilon >1$ can escape from the black hole at a close position $r_{c}$, which is determined by Eqs. \eqref{74} and \eqref{75}, also illustrated in Fig.\ref{Fig:Dcrc}.
 For scattering particles, the radial coordinate is divided into two segments according to whether $r \ge r_{c1}$, where $r_{c1}$ is determined by $\partial_{r} \bar{R}|_{ \varepsilon =1}=0,\bar{R}|_{ \varepsilon =1}=0$, see the intersection point of $r_{c}$ curve with $E_{min}=1$ line in Fig.\ref{Fig:Dcrc}.  In the range $r \ge r_{c1}$, the lower energy of scattering particles is  $\varepsilon_{min }=1$. For those particles which can escape from the black hole in the range $r <r_{c1}$, they must have energy $\varepsilon > \varepsilon_{min}>1$.
When performing the actual calculation, we use FindRoot command to obtain the inverse function $r_c(E_{min})$ appeared in Fig.\ref{Fig:Dcrc}. Therefore, the lower limit of $\varepsilon$ is
 \begin{equation}
   \varepsilon_{min}=\left\{
                       \begin{array}{ll}
                         \varepsilon(r), & \hbox{$r<r_{c1}$;} \\
                         1, & \hbox{$r\geq r_{c1}$.}
                       \end{array}
                     \right.
 \end{equation}
The upper limit of $\varepsilon$ is still infinity. For the limits of $\tau$, the scattering part must have a perihelion, thus $\tau>\tau_c$. At the same time, the signature of function $R$ must be positive. Solving $ \bar{R}=0$,
 one obtains $\bar{\tau}=\bar{\tau}_{max}$. Notice that $\bar{\tau}_{max}$ depends on both $\varepsilon$ and $\sigma$. Thus, the range of $\bar{\tau}$ satisfies $\bar{\tau}_c>\bar{\tau}>\bar{\tau}_{max}$. In addition, we have to note that the trajectory of scattering part contains both $+\sqrt{R}$ and $-\sqrt{R}$, thus
\begin{widetext}
\begin{align}
  J^{scat}_t (r,\theta) &=Am_0^4\sum_{\pm}\int^{\frac{\pi}{2}}_{-\frac{\pi}{2}}d\sigma\int_{\varepsilon_{min}}^{\infty}d\varepsilon\int_{\bar{\tau}_{c}}^{\bar{\tau}_{max}}d\bar{\tau}\frac{\varepsilon e^{-\bar{\beta}\varepsilon}|\cos\theta|}{2\sqrt{\bar{R}}},\label{Jt2}  \\
  J^{scat}_r (r,\theta) &=Am_0^4\sum_{\pm}\int^{\frac{\pi}{2}}_{-\frac{\pi}{2}}d\sigma\int_{\varepsilon_{min}}^{\infty}d\varepsilon\int_{\bar{\tau}_{c}}^{\bar{\tau}_{max}}d\bar{\tau}\frac{ e^{-\bar{\beta}\varepsilon}|\cos\theta|}{2\Delta},\label{Jr2}\\
  J^{scat}_{\theta}(r,\theta)&=Am_0^4\sum_{\pm}\int^{\frac{\pi}{2}}_{-\frac{\pi}{2}}d\sigma\int_{\varepsilon_{min}}^{\infty}d\varepsilon \int_{\bar{\tau}_{c}}^{\bar{\tau}_{max}}d\bar{\tau}\frac{\bar{X} e^{-\bar{\beta}\varepsilon}\cos\sigma\cos^2\theta}{2\sqrt{\bar{R}}\sin\theta},\label{Jtheta2}\\
  J^{scat}_{\varphi}(r,\theta)&=-Am_0^4\sum_{\pm}\int^{\frac{\pi}{2}}_{-\frac{\pi}{2}}d\sigma\int_{\varepsilon_{min}}^{\infty}d\varepsilon\int_{\bar{\tau}_{c}}^{\bar{\tau}_{max}}d\bar{\tau}\frac{\bar{X} e^{-\bar{\beta}\varepsilon}\sin\sigma|\cos\theta|}{2\sqrt{\bar{R}}}.\label{Jphi2}
\end{align}
\end{widetext}

\subsection{Equatorial plane $\theta=\frac{\pi}{2}$}

	The calculation in equatorial plane is very similar to non-equatorial plane. We define
\begin{align}
 L_z& =m_0l_z, \\
 D & =m_0^2\xi,
\end{align}
and
\begin{align}
 \bar{R}_0&=(\varepsilon^2-1)r^4+2Mr^3+(a^2(\varepsilon^2-1)-l_z^2-\xi)r^2+2M(\xi+(l_z-a\varepsilon)^2)r-a^2\xi,
\end{align}
where $R=m_0\bar{R}_0$. $J_{\mu}$ in absorption part are expressed as
\begin{align}
  J^{abs}_t (r,\frac{\pi}{2})&=Am_0^4\int_{-\infty}^{\infty}dl_z\int_1^{\infty}d\varepsilon\int_{0}^{\xi_c}d\xi\frac{\varepsilon e^{-\bar{\beta}\varepsilon}}{2\sqrt{\xi}\sqrt{\bar{R}_0}},\label{Jt01}  \\
  J^{abs}_r(r,\frac{\pi}{2}) &=Am_0^4\int_{-\infty}^{\infty}dl_z\int_1^{\infty}d\varepsilon\int_{0}^{\xi_c}d\xi\frac{ e^{-\bar{\beta}\varepsilon}}{2\Delta\sqrt{\xi}},\label{Jr01}\\
  J^{abs}_{\theta}(r,\frac{\pi}{2})&=Am_0^4\int_{-\infty}^{\infty}dl_z\int_1^{\infty}d\varepsilon\int_{0}^{\xi_c}d\xi\frac{ e^{-\bar{\beta}\varepsilon}}{2\sqrt{\bar{R}_0}},\label{Jtheta01}\\
  J^{abs}_{\varphi}(r,\frac{\pi}{2})&=-Am_0^4\int_{-\infty}^{\infty}dl_z\int_1^{\infty}d\varepsilon\int_{0}^{\xi_c}d\xi\frac{l_z e^{-\bar{\beta}\varepsilon}}{2\sqrt{\xi}\sqrt{\bar{R}_0}},\label{Jphi01}
\end{align}
and in scattering part are expressed as
\begin{align}
  J^{scat}_t (r,\frac{\pi}{2}) &=Am_0^4\sum_{\pm}\int^{\infty}_{-\infty}dl_z\int_{\varepsilon_{min}}^{\infty}d\varepsilon\int_{\xi_{c}}^{\xi_{max}}d\xi\frac{\varepsilon e^{-\bar{\beta}\varepsilon}}{2\sqrt{\xi}\sqrt{\bar{R}_0}},\label{Jt02}  \\
  J^{scat}_r (r,\frac{\pi}{2}) &=Am_0^4\sum_{\pm}\int^{\infty}_{-\infty}dl_z\int_{\varepsilon_{min}}^{\infty}d\varepsilon\int_{\xi_{c}}^{\xi_{max}}d\xi\frac{ e^{-\bar{\beta}\varepsilon}}{2\Delta\sqrt{\xi}},\label{Jr02}\\
  J^{scat}_{\theta}(r,\frac{\pi}{2})&=Am_0^4\sum_{\pm}\int^{\infty}_{-\infty}dl_z\int_{\varepsilon_{min}}^{\infty}d\varepsilon \int_{\xi_{c}}^{\xi_{max}}d\xi\frac{ e^{-\bar{\beta}\varepsilon}}{2\sqrt{\bar{R}_0}},\label{Jtheta02}\\
  J^{scat}_{\varphi}(r,\frac{\pi}{2})&=-Am_0^4\sum_{\pm}\int^{\infty}_{-\infty}dl_z\int_{\varepsilon_{min}}^{\infty}d\varepsilon\int_{\xi_{c}}^{\xi_{max}}d\xi\frac{l_z e^{-\bar{\beta}\varepsilon}}{2\sqrt{\xi}\sqrt{\bar{R}_0}}.\label{Jphi02}
\end{align}
In fact, the values of $J_{\mu}(r,\theta)$ and $T_{\mu\nu}(r,\theta)$ are continuous as $\theta$ increasing from $\theta=0$ to $\theta=\frac{\pi}{2}$.
\section{Asymptotic behavior}
Before doing the numerical computation, we analytically discuss the asymptotic behavior of $J_{\mu}(r,\theta)$ and $T_{\mu\nu}(r,\theta)$. By using Taylor series, the asymptotic behavior can be calculated at large distance. For brevity, we simply define $Am_0^4=Am_0^5=1$ and choose $\bar{\beta}=1$.

	Far away from the black hole, the absorption part becomes zero and only contributions of the scattering part remains. Using Taylor expansion of $\frac{1}{r}$, the function $\frac{1}{\sqrt{\bar{R}}}$ is expressed as
\begin{equation}
  \frac{1}{\sqrt{\bar{R}}}=\frac{1}{\sqrt{\varepsilon^2-1}}\big(\frac{1}{r^2}-\frac{M}{(\varepsilon^2-1)r^3}\big),
\end{equation}
where only the first two terms are listed. Since the integrand is independent of $\bar{\tau}$, the integral with respect to $\bar{\tau}$ simply is
\begin{equation}
  \int_{\bar{\tau}_c}^{\bar{\tau}_{max}}d\bar{\tau}=\bar{\tau}_{max}-\bar{\tau}_c,
\end{equation}
where $\bar{\tau}_c$ does not depend on $r$. Since $\bar{\tau}_{max}$ is determined by $\bar{R}=0$, it is expressed as
\begin{equation}
 \bar{\tau}_{max}=\frac{r^2}{\cos^2\theta+\sin^2\theta\sin^2\sigma}\left((\varepsilon^2-1)+2M\varepsilon^2 \frac{1}{r}\right)
\end{equation}
using Taylor expansion. The direct calculation of equation (\ref{Jt2}) obtains
\begin{align}
  J_{t}(r\rightarrow\infty )&=\int_{-\frac{\pi}{2}}^{\frac{\pi}{2}}d\sigma\int_1^{\infty}d\varepsilon\frac{\bar{\tau}_{max} }{\sqrt{\bar{R}}}\varepsilon e^{-\varepsilon}|\cos\theta|\nonumber\\
&=\pi\left(K_{2}(1)+(2K_{0}(1)+5K_{1}(1))\frac{M}{r}\right),
\end{align}
where $K_{n}(x)$ is the second Bessel function. The values of $K_{n}(x)$ can be found in the related books. Thus, the above expression approximates to
\begin{equation}
  J_{t}(r\rightarrow\infty)=5.104+12.100\frac{M}{r},
\end{equation}
which agrees with the numerical computation in the next section. Notice that the expression of $J_{t}$ is independent of $\theta$ at infinity. If $\bar{\beta}$ is an arbitrary constant, the first term of $J_t(r\rightarrow\infty)$ is
\begin{equation}
  J_t(r\rightarrow\infty)= \frac{\pi K_{2}(\bar{\beta})}{\bar{\beta}}.
\end{equation}
Thus, the particle density $n_{\infty}$ at infinity is
\begin{equation}
  n_{\infty}=\pi\frac{ K_{2}(\bar{\beta})}{\bar{\beta}}.
\end{equation}
This result is proportional to that in Ref. \cite{Rioseco2017-1}, also see the Appendix B in Ref. \cite{Cieslik2022}. The difference can be eliminated by choosing the normalized constant $A$ appropriately.

	Similar to the calculation of $J_{t}(r\rightarrow\infty)$, the values of other components at infinity can be calculated as follows
\begin{align}
  J_r(r\rightarrow\infty) &=\frac{4\pi}{e}+\frac{18\pi}{e}\frac{M}{r},  \\
T_{tr}(r\rightarrow\infty) & =\frac{14\pi}{e}+\frac{60\pi}{e}\frac{M}{r},\\
T_{r\varphi }(r\rightarrow\infty) & =0.
\end{align}
At large distance, Kerr metric approximates to Schwarzschild one and the parameter $a$ has no effect. This is the main reason that $J_{\mu}$ and $T_{\mu\nu}$ are independent of $\theta$ at infinity.
\begin{figure}[htb]
\centering
\includegraphics[width=0.45\textwidth]{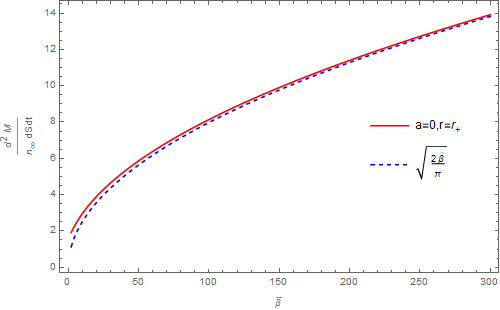}
\caption{The asymptotic behavior of unit mass accretion rate in low temperature limit. As $\bar{\beta}$ increasing, the unit mass accretion rate coincides with the function $\sqrt{\frac{2\bar{\beta}}{\pi}}$.}
\label{Fig:Asy}
\end{figure}

	Our results can approximate to the Schwarzschild case in low temperature limit. The process is as follows. Near the horizon $r_{+}$, the scattering part is omitted. In the special case $a=0$, the background spacetime approximates to Schwarzschild metric. The critical value of $\bar{\tau}_c$ can be calculated as
\begin{widetext}
\begin{equation}
\bar{\tau}_c=\frac{ M^2\left(9 \left(3 \epsilon ^2+\sqrt{\epsilon ^2 \left(9 \epsilon ^2-8\right)}-4\right) \epsilon ^2-8 \sqrt{\epsilon ^2 \left(9 \epsilon ^2-8\right)}+8\right)}{2 \left(\epsilon ^2-1\right) \left(\sin ^2\theta  \sin ^2\sigma +\cos ^2\theta \right)}.
\end{equation}
\end{widetext}
We notice that as $r\rightarrow r_{+}$, $\Delta(r_{+})$ turns into $0$ and $J_{r}$ becomes infinity. However, the unit accretion mass $\frac{d^2M}{dS dt}$ is a finite value. We obtain
\begin{widetext}
\begin{equation}\label{AM1}
 \frac{d^2M}{n_{\infty}dSdt}\bigg|_{r_+}=\frac{\bar{\beta}}{16 K_2(\bar{\beta})}\int_1^{\infty}\frac{ 9 \left(3 \epsilon ^2+\sqrt{\epsilon ^2 \left(9 \epsilon ^2-8\right)}-4\right) \epsilon ^2-8 \sqrt{\epsilon ^2 \left(9 \epsilon ^2-8\right)}+8}{ \left(\epsilon ^2-1\right) }e^{-\bar{\beta} \varepsilon} d\varepsilon.
\end{equation}
\end{widetext}
In the low temperature limit $\bar{\beta}\rightarrow\infty$, Ref. \cite{Rioseco2017-1} gave a result of mass accretion rate
\begin{equation}
  \frac{1}{n_{\infty}}\frac{\mathrm{d}M}{\mathrm{d}t}\bigg|_{low}=16\sqrt{2\pi}M^2\sqrt{\bar{\beta}},
\end{equation}
where compared to Ref. [12], the difference is just due to the normalization we chose that $m_0=1$. Considering the area of horizon surface $S=4\pi r_{+}^2=16\pi M^2$, we have
\begin{equation}\label{AM2}
  \frac{d^2M}{n_{\infty}dSdt}\bigg|_{low}=\sqrt{\frac{2\bar{\beta}}{\pi}}.
\end{equation}
The numerical calculation shows that Eq. (\ref{AM1}) and Eq. (\ref{AM2}) coincide with each other as $\bar{\beta}$ increasing, see Fig.\ref{Fig:Asy}.

\section{Numerical results}
\begin{figure}[htbp]
\centering
\includegraphics[width=0.45\textwidth]{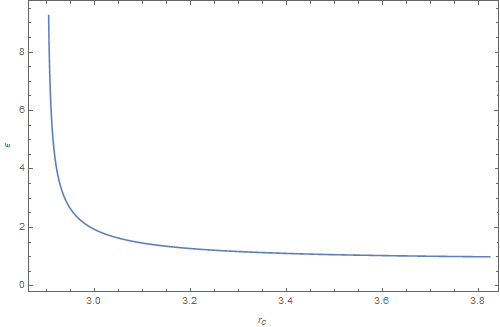}
\caption{The numerical relationship between $\varepsilon$ and $r_c$. The minimum value of $r_c$ approaches the photon sphere of the Kerr black hole. Parameters are chosen to be $\theta=\frac{\pi}{3},\sigma=\frac{\pi}{3},a=0.1$.} \label{Fig:Eprc}
\end{figure}

In doing the numerical computation, without loss of generality, we choose $M=\bar{\beta}=m_{0}=A=1$.  Similar to Fig.\ref{Fig:Dcrc}, the numerical relationship between $\varepsilon$ and $r_c$ is plotted in Fig.\ref{Fig:Eprc}. In the calculation of scattering part, function $\varepsilon(r_c)$ determines the lower integral bound of $\varepsilon$. The common method is to find the inverse function of $r_c(\varepsilon)$ through FindRoot.

\begin{figure}[htbp]
\centering
\includegraphics[width=0.45\textwidth]{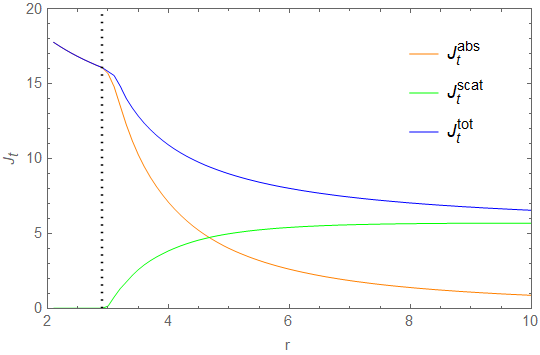}
\caption{$J_t$ as a function of $r$. Parameters are chosen to be $\theta=\frac{\pi}{3},a=0.1$. The upper integral bound of $\varepsilon$ is truncated to $\varepsilon=8$. The dot line stands for photon sphere $r=r_{ph}$. In the range $r<r_{ph}$, there are no scattered particles.}
\label{Fig:Jt}
\end{figure}

\begin{figure}[htbp]
\centering
\includegraphics[width=0.45\textwidth]{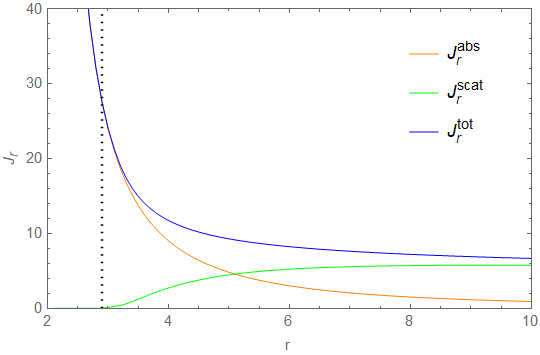}
\caption{$J_r$ as a function of $r$. Parameters are chosen to be $\theta=\frac{\pi}{3},a=0.1$.}
\label{Fig:Jr}
\end{figure}

\begin{figure}[htbp]
\centering
\includegraphics[width=0.45\textwidth]{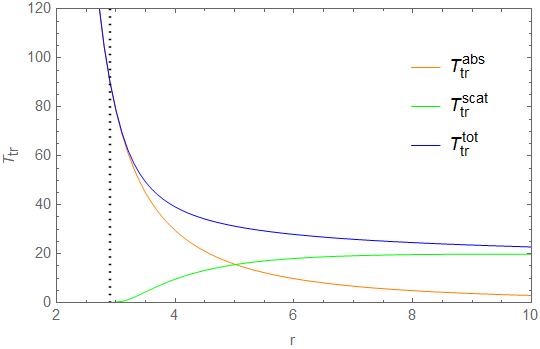}
\caption{$T_{tr}$ as a function of $r$. Parameters are chosen to be $\theta=\frac{\pi}{3},a=0.1$.}
\label{Fig:Ttr}
\end{figure}

\begin{figure}[htbp]
\centering
\includegraphics[width=0.45\textwidth]{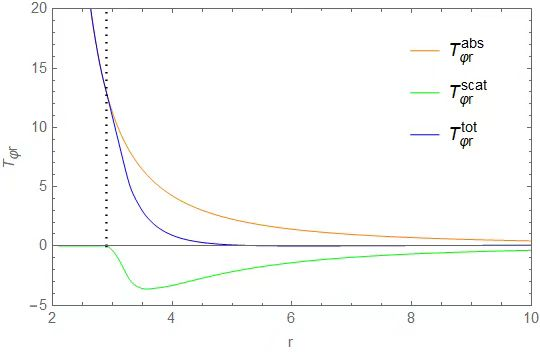}
\caption{$T_{\varphi r}$ as a function of $r$. Parameters are chosen to be $\theta=\frac{\pi}{3},a=0.1$.}
\label{Fig:Tphir}
\end{figure}

\begin{figure}[htbp]
\centering
\includegraphics[width=0.45\textwidth]{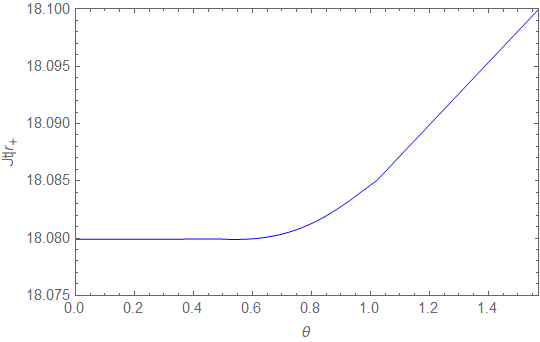}
\caption{$J_{t}|_{r_{+}}$ as a function of $\theta$. Parameter is chosen to be $a=0.1$. It increases slightly when $\theta$ changes from 0 to $\frac{\pi}{2}$.}
\label{Fig:Jttheta}
\end{figure}

\begin{figure}[htbp]
\centering
\includegraphics[width=0.55\textwidth]{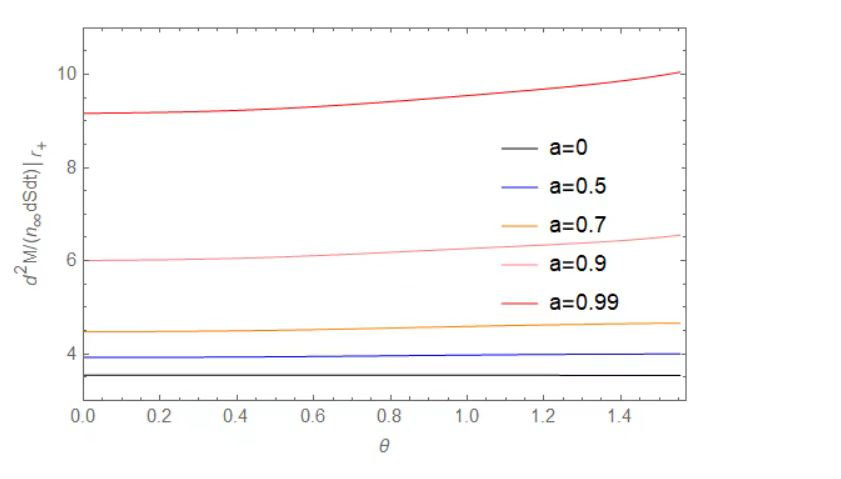}
\caption{Unit mass accretion rate $\frac{d^2M}{n_{\infty}dSdt}\bigg|_{r_+}$ as a function of $\theta$. The range of $\theta$ is throughout the Northern Hemisphere $\theta\in[0,\frac{\pi}{2}]$. The behavior in the Southern Hemisphere is symmetric to that in the northern hemisphere.}
\label{Fig:Mass}
\end{figure}

	The variations of $J_t$ with $r$ are plotted in Fig.\ref{Fig:Jt}, which shows that with $r$ increasing, the absorption part is monotonically decreasing, and the scattering part and the total are monotonically increasing. $J_{r}, T_{tr}$ and $T_{\varphi r}$ have similar behaviors shown in Fig.\ref{Fig:Jr}, Fig.\ref{Fig:Ttr}, and Fig.\ref{Fig:Tphir}, respectively.

	We also consider the variations of the particle current densities and the stress energy momentum with respect to $\theta$ at the horizon. Since $J_{r}, T_{tr}$ and $T_{\varphi r}$ approach to infinity at the horizon, we plot $J_{t}|_{r_{+}}$ as a function of $\theta$ in Fig.\ref{Fig:Jttheta}. It is shown that $J_{t}|_{r_{+}}$ increases slightly when $\theta$ changes from 0 to $\frac{\pi}{2}$.

\begin{figure}[htbp]
\centering
\includegraphics[width=0.45\textwidth]{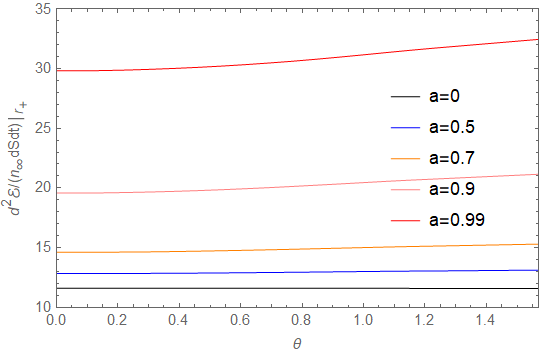}
\caption{Unit energy accretion rate $\frac{d^2\mathcal{E}}{n_{\infty}dSdt}\bigg|_{r_+}$ as a function of $\theta$.}
\label{Fig:Energy}
\end{figure}

\begin{figure}[htbp]
\centering
\includegraphics[width=0.45\textwidth]{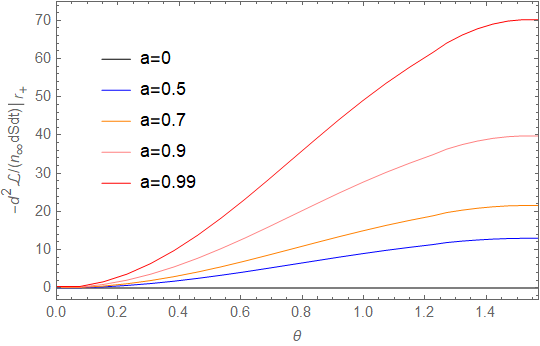}
\caption{Unit angular momentum accretion rate $-\frac{d^2\mathcal{L}}{n_{\infty}dSdt}\bigg|_{r_+}$ as a function of $\theta$.}
\label{Fig:Momentum}
\end{figure}

	Furthermore, the unit accretion rates of the black hole are much relevant in physics. In this paper, we concern about the accretion rates changing with $\theta$ at the horizon. We plot the unit mass accretion rate $\frac{d^2M}{n_{\infty}dSdt}\bigg|_{r_+}$ as a function of $\theta$ in Fig.\ref{Fig:Mass}, which shows a weak growth as $\theta$ approaches the equatorial plane. A similar behavior also appears in energy accretion rate $\frac{d^2\mathcal{E}}{n_{\infty}dSdt}\bigg|_{r_+}$, see Fig.\ref{Fig:Energy}. However, as seen in Fig.\ref{Fig:Momentum}, the unit momentum accretion rate grows more rapidly as $\theta$ increasing. These behaviors imply that the unit accretion rates at the horizon surface have different growth trends. As the accretion goes on, the background would deviate from the Kerr one, depending on the total mass of particles falling into the black hole. Fig.\ref{Fig:Mass} also shows that the unit mass accretion rate increases nonlinearly as $a$ grows, which would further alleviate the BHL accretion problem. From Fig.\ref{Fig:Momentum}, we also know that for larger $a$, the black hole is slowed down more quickly by accretion.
	
\section{Conclusion}

	In this paper, we study the accretion of Vlasov gas onto a Kerr black hole, where particles are distributed throughout all the spacetime. The main conclusions are as follows.

	We solve the the relativistic Liouville equation in Kerr background.  We prove that the distribution function is independent of $Q^{\mu}$ if the flow is stationary and axisymmetric.

	We derive the expression of the volume element corresponding to equatorial plane and non-equatorial plane. We also obtain the expressions of three unit accretion rates surrounded by a sphere.

	In an example of Maxwell-J\"{u}ttner distribution at infinity, we further calculate the particle current density $J_{\mu}$, the stress energy-momentum tensor $T_{\mu\nu}$ and three unit accretion rates. At large distance, we first give a convenient analytical computing method, which shows all quantities are independent on $\theta$. And the correlative results are approximate to those in the Schwarzschild case. In the non-rotating case $a\rightarrow 0$, the unit mass accretion rate is computed by numerical method and coincident to the result in Schwarzschild case. In the finite range, the corresponding calculation results are obtained by numerical method.

The numerical results show growths of accretion rates as $\theta$ approaches the equatorial plane. If the mass of particles falling into the black hole is much less than the mass of the black hole, that is, $m_{gas}<<M$, the background metric remains unchanged. However, as the Vlasov particles keep flowing into the black hole, $m_{gas}$ increases continuously, and the background metric may change inevitably.

	The numerical results also show nonlinear growth of the accretion rates as $a$ approaches to $M$, which implies the possibility of allevicating the BHL accretion problem if the black hole is rotating.

  As is known, a realistic accretion flow will involve magnetic fields, which will mediate effective collisionality between the particles. To consider the effect of magnetic fields will be an interesting topic worth studying in the future.

\appendix
\section{ The components of $T_{\mu\nu}$}
The components of $T_{\mu\nu}$ are
 \begin{align}
   T^{i}_{tt}(r,\theta) &=Am_0^5 \iiint_{V_i} \frac{\varepsilon^2 e^{-\bar{\beta}\varepsilon}|\cos\theta|}{2\sqrt{\bar{R}}}d\sigma d\varepsilon d\bar{\tau}, \\
   T^{i}_{tr}(r,\theta) &=Am_0^5 \iiint_{V_i} \frac{\varepsilon e^{-\bar{\beta}\varepsilon}|\cos\theta|}{2\Delta}d\sigma d\varepsilon d\bar{\tau}, \label{Ttr1}\\
   T^{i}_{t\theta}(r,\theta) &=Am_0^5 \iiint_{V_i} \frac{\bar{X}\varepsilon e^{-\bar{\beta}\varepsilon}\cos\sigma\cos^2\theta}{2\sqrt{\bar{R}}\sin\theta}d\sigma d\varepsilon d\bar{\tau},\label{Tttheta1} \\
 T^{i}_{t\varphi}(r,\theta) &=-Am_0^5 \iiint_{V_i} \frac{\varepsilon\bar{X} e^{-\bar{\beta}\varepsilon}\sin\sigma|\cos\theta|}{2\sqrt{\bar{R}}}d\sigma d\varepsilon d\bar{\tau}, \\
 T^{i}_{rr}(r,\theta) &=Am_0^5 \iiint_{V_i} \frac{\sqrt{\bar{R}} e^{-\bar{\beta}\varepsilon}|\cos\theta|}{2\Delta^2}d\sigma d\varepsilon d\bar{\tau}, \\
 T^{i}_{r\theta}(r,\theta) &=Am_0^5 \iiint_{V_i} \frac{\bar{X} e^{-\bar{\beta}\varepsilon}\cos\sigma\cos\theta^2}{2\Delta\sin\theta}d\sigma d\varepsilon d\bar{\tau}, \\
   T^{i}_{r\varphi }(r,\theta)&=-Am_0^5 \iiint_{V_i}\frac{\bar{X} e^{-\bar{\beta}\varepsilon}\sin\sigma\cos\theta}{2\Delta}d\sigma d\varepsilon d\bar{\tau},\label{Tphir1}\\
 T^{i}_{\theta\theta}(r,\theta) &=Am_0^5 \iiint_{V_i} \frac{\bar{X}^2 e^{-\bar{\beta}\varepsilon}\cos^2\sigma|\cos^3\theta|}{2\sqrt{\bar{R}}\sin^2\theta}d\sigma d\varepsilon d\bar{\tau}, \\
   T^{i}_{ \theta \varphi}(r,\theta)&=-Am_0^5 \iiint_{V_i}\frac{\bar{X}^2 e^{-\bar{\beta}\varepsilon}\sin\sigma\cos\sigma\cos^2\theta}{2\sqrt{\bar{R}}\sin\theta}d\sigma d\varepsilon d\bar{\tau},\label{Tphitheta1}\\
 T^{i}_{\varphi\varphi}(r,\theta) &=Am_0^5 \iiint_{V_i} \frac{\bar{X}^2 e^{-\bar{\beta}\varepsilon}\sin^2\sigma|\cos\theta|}{2\sqrt{\bar{R}}}d\sigma d\varepsilon d\bar{\tau},
 \end{align}
where $i$ stands for absorption or scattering. The domain $V_i$ of integration is determined by the case of absorption or scattering. For absorption part, the interval are $\sigma\in[-\frac{\pi}{2},\frac{\pi}{2}],\varepsilon\in[1,\infty),\bar{\tau}\in[0,\bar{\tau}_c]$. For scattering part, the interval are $\sigma\in[-\frac{\pi}{2},\frac{\pi}{2}],\varepsilon\in[\varepsilon_{min},\infty),\bar{\tau}\in[\bar{\tau}_c,\bar{\tau}_{max}]$.

\nocite{*}

\bibliography{apssamp}% Produces the bibliography via BibTeX.

\end{document}